# Metal-nanoparticle single-electron transistors fabricated using electromigration

K. I. Bolotin[*], F. Kuemmeth[*], A. N. Pasupathy, and D. C. Ralph
*Laboratory of Atomic and Solid State Physics, Cornell University, Ithaca NY, 14853*
[*] These authors contributed equally to this work

(Dated October 9, 2003)

We have fabricated single-electron transistors from individual metal nanoparticles using a geometry that provides improved coupling between the particle and the gate electrode. This is accomplished by incorporating a nanoparticle into a gap created between two electrodes using electromigration, all on top of an oxidized aluminum gate. We achieve sufficient gate coupling to access more than ten charge states of individual gold nanoparticles (5-15 nm in diameter). The devices are sufficiently stable to permit spectroscopic studies of the electron-in-a-box level spectra within the nanoparticle as its charge state is varied.

Single-electron transistors (SETs) made from metals are important for their use as electrometers, and also because they enable fundamental studies of charge transport in metals.[1-8] For many applications, it is desirable to minimize the size of the central metal island within the SET, so as to increase the charging energy, optimize the electrometer sensitivity or spatial resolution, or raise the maximum operating temperature. If the island is made smaller than about 10 nm diameter, it is also possible to use such SETs to probe individual "electron-in-a-box" quantum states within the metal nanoparticle.[8] SETs with metal islands smaller than 10 nm have been fabricated previously,[9,10] but the gate capacitances were so small ($10^{-19}$ F in ref. 9 and $< 10^{-20}$ F in ref. 10) that it was difficult to adjust the charge on the island by more than one electron. Here we describe a technique to make metal-nanoparticle SETs with gate capacitances of order $10^{-18}$ F, sufficient to tune the number of electrons in a nanoparticle by more than ten. We use these devices to study the evolution of the electron-in-a-box level spectrum in gold nanoparticles as the electron number is changed one by one.

Our fabrication technique builds upon work in which electrical contact was made to single nanoparticles without a gate electrode[10-13] and to networks of particles in the presence of a gate.[14] We start by using photolithography to define a 16-nm-thick Al gate electrode with 2 nm of Ti as a sticking layer, on top of an oxidized Si substrate. The gate electrode is deposited with the substrate at liquid nitrogen temperature. The Al is warmed overnight to room temperature while in 50 mtorr of $O_2$ and then exposed to air.[15] Next we use electron-beam lithography and liftoff to fabricate Au wires with a thickness of 16 nm and a minimum width of 100 nm on top of the gate (Fig. 1a). After cleaning the Au wires in oxygen plasma, we submerge the chip in liquid helium and break the wires using electromigration;[16] a source-drain bias is slowly ramped up until the wire breaks and the conductance drops suddenly.[17] In most cases this happens at a bias of ~1 V, and results in a gap about 5-10 nm wide after the sample is warmed to room temperature (Fig. 1b). A similar process of electromigration has been used recently to make single-molecule transistors.[18,19] Then we create nanoparticles, formed by self-assembly, by evaporating Au on top of the broken lines and into the gap with the substrate at room temperature. For most samples we evaporate 20 Å of gold at 0.6 Å/s, which produces particle diameters of 5-15 nm as measured with SEM (Fig. 1b). When measured immediately after the evaporation, approximately 25% of the junctions show a drop of resistance into the MΩ range, indicating that a nanoparticle may be bridging the gap. This procedure has the virtue of producing nanoparticles separated from the gate electrode by only 2-3 nm of aluminum oxide, giving excellent coupling between the particle and the gate. Finally the devices are protected by evaporating 100 nm of aluminum oxide prior to air exposure.

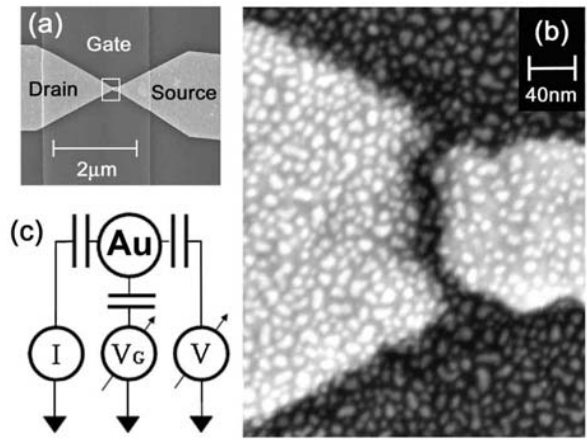

FIG. 1. (a) Top-view scanning electron microscope (SEM) image of the device geometry, with gold source and drain electrodes on top of an oxidized aluminum gate. (b) Expanded view of the region outlined with a white rectangle in (a). A 10-nm gap made by electromigration is visible, along with deposited gold nanoparticles. (c) Circuit schematic for the Au-nanoparticle SET.

We cool the finished devices to 4.2 K and measure their current-voltage (*I-V*) curves as a function of gate voltage ($V_G$) (see Fig. 1c for circuit conventions). More than 50% of the devices that show a decrease in resistance during the nanoparticle deposition step exhibit Coulomb-blockade characteristics (Fig. 2a and 2b). Fits to the orthodox theory of Coulomb blockade[20] yield good agreement (Fig. 2a). The variation of $dI/dV$ as a function of $V_G$ and $V$ is shown in Fig. 2c, for a range of $V_G$ large enough to span several "Coulomb diamonds". The presence of only one positive and one negative slope for the tunneling thresholds vs. $V_G$ indicates that transport is indeed occurring through a single nanoparticle.[21] The gate capacitance $C_G$ is determined from the spacing in $V_G$ between the degeneracy points where $dI/dV$ is nonzero at $V=0$. For device #1 (Fig. 2a-c), the spacing between neighboring degeneracy points is 180 mV, so $C_G = e/(180\text{ mV}) = 0.89$ aF ($\pm 5\%$). This is an order of magnitude greater than gate capacitances achieved previously in metal nanoparticle transistors.[9] The ratios $C_D/C_G$ and $C_S/C_G$ for the drain and source capacitances are then determined from the slopes of the tunneling thresholds vs. $V_G$,[22] yielding for device #1 $C_D = 2.4$ aF and $C_S = 1.3$ aF ($\pm 10\%$). Fig. 2c demonstrates that 12 different charge states within this nanoparticle can be accessed by varying $V_G$ within the range $\pm 1$ V. We should note that the scan in Fig. 2c is somewhat atypical in that there are very few discontinuities in the gate-voltage dependence, corresponding to sudden rearrangements in the background charge near the particle. The diamond plot in Fig. 2d, from device #2, is more typical.

The yield given by our fabrication process is that 15-20% of all devices exhibit gate-dependent Coulomb blockade. The gate capacitances are typically 0.3-2.0 aF and the source and drain capacitances are 1-10 times larger, yielding charging energies between 15 and 50 meV. As a control experiment, we took 24 devices through the full fabrication process except for the step in which the nanoparticles are deposited. These samples never showed Coulomb-blockade characteristics.

In prior work, tunnel junctions containing individual metal nanoparticles have been used to measure the "electron-in-a-box" states inside the nanoparticle, and these experiments have been a rich source of information about electronic interactions within metals.[8,9] Having a more effective gate would contribute greatly to these types of studies, by enabling investigations of how the spectra change as electrons are added to a nanoparticle, and also by giving increased control over non-equilibrium excitations.[22] Next we demonstrate that our SETs are sufficiently stable to allow measurements of the electron-in-a-box spectra, and we present initial studies of how these spectra depend on the electron number.

In order to resolve electron-in-a-box quantum states,

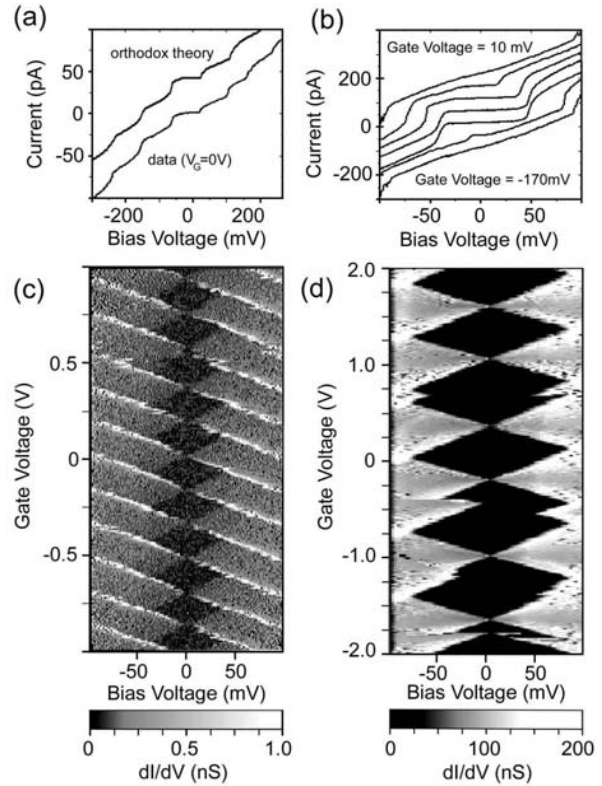

FIG. 2. (a) Coulomb staircase $I$-$V$ curve for a gold nanoparticle SET (device #1) at 4.2 K, along with an orthodox model fit (offset for clarity). (b) $I$-$V$ curves for device #1 at 4.2 K, for equally spaced values of $V_G$. (c) Gray-scale plot of $dI/dV$ as a function of $V_G$ and $V$ for device #1 at 4.2 K. Eleven degeneracy points separating twelve different charge states are visible within a 2 V range of $V_G$. (d) Gray-scale plot of $dI/dV$ as a function of $V_G$ and $V$ for device #2 at 4.2 K, showing "Coulomb diamonds" as well as several abrupt changes in the background charge of the SET island as $V_G$ is swept.

we cool samples in a dilution refrigerator with filtered electrical lines so that the electron temperature is less than 150 mK. In Fig. 3 we examine a device with parameters $C_G = 2.0$ aF, $C_S = 1.9$ aF and $C_D = 2.2$ aF. Low-temperature plots of $dI/dV$ versus $V_G$ and $V$ reveal a fine structure of lines beyond the tunneling threshold that correspond to tunneling via excited quantum states within the nanoparticle. Figure 3 shows three sets of spectra. In panel (a), the lines running parallel to the line labeled α correspond to transitions in which an electron tunnels off the nanoparticle to decrease the total number of electrons from some value N to N-1. Panel (b) displays a more negative range of $V_G$ where transitions from N-1 electrons to N-2 are visible. Finally, the levels in panel (c) correspond to N-2 to N-3 transitions. The most striking aspect of these three plots is that the pattern of excited states is extremely similar. The only significant difference can be seen in the line labeled α at the threshold for tunneling. In panel (a) line α is strong, in panel (b) it is present but its conductance is decreased by approximately 45%, and in (c) the line is absent. This behavior can be understood in a simple way by assuming that the energy levels in the nanoparticle are spin-degenerate, and that N corresponds to an even number of electrons. In this case, line α in panel (a) corresponds to transitions in which either one of the two (spin-degenerate) electrons in the highest occupied energy level for N electrons tunnels off of the particle into the source. For N-1 electrons in panel (b), one of these two electrons is already gone, so line α corresponds to having just the one remaining electron in that same energy level tunnel off of the particle. Because the source-particle interface is the rate limiting barrier in this device, the tunneling rate is reduced by approximately a factor of 2. Finally, in panel (c), no electrons are left to tunnel out of the state corresponding to line α, so the line is no longer present. The strong similarities that we observe in the excited state spectra for different numbers of electrons on a gold nanoparticle are in striking contrast to analogous studies of GaAs quantum dots.[23] In GaAs dots, spin degeneracy is not generally observed and excited-state levels are shifted relative to one another when the electron number is changed. These spectral rearrangements in GaAs have been explained as due to exchange interactions between the added electron and quantum-dot states with different total spins. The similarity of the excited-state spectra that we observe for different numbers of electrons in gold is therefore evidence that exchange interactions are sufficiently weak that the energy levels are filled as in a non-interacting model, with each energy level spin-degenerate and with the energy-level spacings not sensitive to the electron number. This simple outcome is highly non-trivial for an interacting-electron system, but it is consistent with expectations for the strength of exchange interactions in gold.[24]

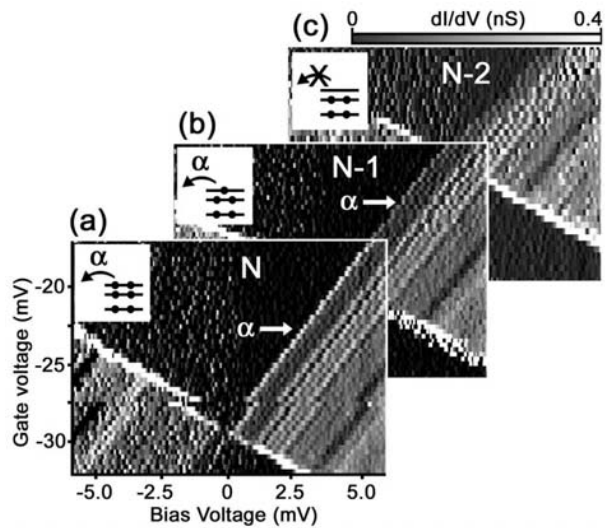

FIG. 3. Gray-scale plots of $dI/dV$ as a function of $V_G$ and $V$, displaying the discrete electron-in-a-box level spectra within a gold nanoparticle (device #3), measured with an electron temperature less than 150 mK and zero magnetic field. Panels (a), (b), and (c) represent spectra for different numbers of electrons in the same nanoparticle. Panel (b) covers the gate voltage range from -95 mV to -110 mV and (c) the range from -180 mV to -195 mV. Insets: Energy-level diagrams illustrating the tunneling transitions that contribute to line α for different numbers of electrons on the particle.

In summary, we have fabricated single-electron transistors by depositing metal nanoparticles within break junctions made using electromigration. This method provides strong coupling between the nanoparticle and the gate electrode, thereby enabling experiments in which the number of electrons on the particle is varied over a wide range. We have demonstrated that this device geometry permits detailed measurements of electron-in-a-box level spectra in metals as a function of the electron number.

We thank A. R. Champagne and J. E. Grose for helpful discussions. This work was funded by the NSF (DMR-0244713 and through use of the Cornell Nanoscale Facility/NNUN) and by the ARO (DAAD19-01-1-0541).